\documentstyle[11pt,aaspp4]{article}

\def\arcs{\ifmmode {^{\scriptscriptstyle\prime\prime}}
          \else $^{\scriptscriptstyle\prime\prime}$\fi}
\def\arcm{\ifmmode {^{\scriptscriptstyle\prime}}
          \else $^{\scriptscriptstyle\prime}$\fi}
\newdimen\sa  \newdimen\sb
\def\parcs{\sa=.07em \sb=.03em
     \ifmmode $\rlap{.}$^{\scriptscriptstyle\prime\kern -\sb\prime}$\kern -\sa$
     \else \rlap{.}$^{\scriptscriptstyle\prime\kern -\sb\prime}$\kern -\sa\fi}
\def\parcm{\sa=.08em \sb=.03em
     \ifmmode $\rlap{.}\kern\sa$^{\scriptscriptstyle\prime}$\kern-\sb$
     \else \rlap{.}\kern\sa$^{\scriptscriptstyle\prime}$\kern-\sb\fi}
\def\pdeg{\ifmmode $\setbox0=\hbox{$^{\circ}$}\rlap{\hskip.11\wd0 .}$^{\circ}
          \else \setbox0=\hbox{$^{\circ}$}\rlap{\hskip.11\wd0 .}$^{\circ}$\fi}
\def\gtorder{\mathrel{\raise.3ex\hbox{$>$}\mkern-14mu
             \lower0.6ex\hbox{$\sim$}}}
\def\ltorder{\mathrel{\raise.3ex\hbox{$<$}\mkern-14mu
             \lower0.6ex\hbox{$\sim$}}}
\def\vs{\vec{s}}

\newcommand\etal{et~al.}

\input psfig
\lefthead{}
\righthead{}

\begin{document}

\title{Evidence for Inconsistencies in \\
  Galaxy Luminosity Functions Defined by Spectral Type\footnote{This publication 
  makes use of data products from the Two Micron All Sky Survey (2MASS), which is a joint project of the 
  University of Massachusetts and the Infrared Processing and Analysis Center/California Institute of 
  Technology, funded by the National Aeronautics and Space Administration and the National Science 
  Foundation. }\footnote{This research has made use of the NASA/IPAC Extragalactic Database (NED) which 
  is operated by the Jet Propulsion Laboratory, California Institute of Technology, under
  contract with the National Aeronautics and Space Administration.}}

\author{C. S. Kochanek, M. A. Pahre\footnote{Hubble Fellow.}, \& E. E. Falco}
\affil{Harvard-Smithsonian Center for Astrophysics \\
       60 Garden Street \\
       Cambridge, MA 02138 \\
       ckochanek, mpahre, efalco@cfa }

\begin{abstract}
Galaxy morphological and spectroscopic types should be nearly
independent of apparent magnitude in a local, magnitude-limited sample.
Recent luminosity function surveys based on morphological
classification of galaxies are substantially more successful at
passing this test than surveys based on spectroscopic
classifications.  Among spectroscopic classifiers, those defined by
small aperture fibers (ESP, LCRS) show far stronger systematic
classification biases than those defined by  large apertures (APM).
This effect can be easily explained as an \emph{aperture bias},
whereby galaxies with globally late-type spectra are assigned earlier
spectral types which depend on the redshift and luminosity of the
galaxy.   The effect is demonstrated by extracting successively larger
aperture spectra from long-slit spectroscopy of nearby galaxies.  
This systematic classification bias is generic, and the 2dFGRS  and
SDSS surveys will show similar systematic biases in their spectral
types.  We suggest several methods for correcting the problem or
avoiding it altogether.  If not corrected, this aperture bias can mimic
galaxy evolutionary effects and distort estimates of the luminosity function.
\end{abstract}


\section{Introduction}

The classification of galaxies into different types is more than mere
phenomenology: it is a method to separate galaxies in a
well-defined manner in order to study their common properties locally and
their evolution with redshift.  The applications of
galaxy classification are quite diverse, since significant differences 
in measurable physical properties can be attributed to various galaxy types:
luminosity functions for distinct galaxy types evolve differently with redshift
(\cite{lilly95}); galaxies differ in their clustering properties
through the  morphology-density relation (\cite{dressler80}) and their
bias with respect to the  underlying mass distribution
(\cite{hubble36}; \cite{oemler74}; \cite{blanton99});  and the
luminosity evolution of early-type (\cite{dokkum96}, \cite{kochanek00b}) and late-type
(\cite{vogt96}) galaxies at fixed mass is also very different for $0<z<1$
due to their different stellar populations.  All these properties
are relevant for estimates of the local galaxy power spectrum 
(\cite{kolatt97}) and for estimates of the cosmological model using
gravitational lenses (e.g. \cite{kochanek96}).

Galaxies can be classified based on morphology, spectra, colors, and
surface photometry to produce a one-dimensional sequence.  While
spectral and morphological methods, for example, yield similar
classification information (\cite{morgan57}), these two methods are
not identical (\cite{connolly95}).  Each method emphasizes a different
combination of physical properties such as current star formation and
AGN activity (emission lines, color), past star formation (absorption lines, 
color), disk instabilities (spiral structure), and merger histories
(bulge-to-disk ratios).  The continuous one-dimensional sequence is
then broken into discrete types at somewhat arbitrary
boundaries. Finally, galaxies are assigned specific types, usually
without uncertainty estimates, and the results are analyzed.  Thus,
while the classification of galaxies is necessary to studies of the
structure, formation, and evolution of galaxies, the differences and
uncertainties in classification methods create many problems of
interpretation.

Morphological classification has a long history dating back to Hubble (1936\markcite{hubble36}),
and has several ``picture books'' which are typically referenced to create the operational 
definition of the morphological types---the ``Hubble Atlas of Galaxies'' (\cite{hubbleatlas}) 
and the Revised Shapley-Ames Catalog (\cite{rsa}).  The method has had a recent
renaissance with its application to galaxies at cosmological distances using the high 
spatial resolution and sensitivity of the Hubble Space Telescope (e.g., field galaxies in the HDF,
\cite{vandenbergh96}; clusters galaxies, \cite{dressler97}).
Spectroscopic classification of galaxies is a more recent field
(\cite{connolly95}) which offers the promise of objective
decomposition of spectra into various galactic components (nucleus,
bulge, and disk), but may suffer from the reduced spatial information
typically available for each galaxy.  While it might be simple to
equate morphological and spectroscopic classifications, their precise
relationship to each other will also be some function of the
observational limitations, such as spatial resolution, rest-frame
wavelength of the filter, spatial extent used to extract the spectra,
etc.  It is still an open question as to which
classification method---morphological, spectroscopic, bulge-to-disk, color,
and so on---has superior precision, reliability, and physical
insight.

In Kochanek \etal\ (2000\markcite{kochanek00a}) we derived the first
local infrared galaxy luminosity function large enough to be compared
to optical galaxy luminosity functions.  As part of our analysis we
morphologically classified the galaxies because comparisons of the
infrared and optical results have to be done by galaxy type due to the
large differences in the $(B-K)$ or $(R-K)$ colors of galaxies of
different morphological or spectral types.  We used morphological
classification because it was relatively easy to do for our bright,
local galaxy sample and because we relied so heavily on archival
redshifts that we lacked the data necessary to attempt spectral
classification.  We could compare our results to eight relatively
recent derivations of luminosity functions divided by galaxy type.
The APM survey has derived results using both morphological
(\cite{loveday92}) and spectral classification (\cite{loveday99}).
The CfA (\cite{marzke94b}) and SSRS2 (\cite{marzke98}) surveys used
morphological types.   The LCRS (\cite{lin96}), ESP (\cite{zucca97}),
and APM (\cite{loveday99}) surveys used emission line equivalent
widths to define types, and the LCRS (\cite{bromley98a}) and 2dFGRS
(\cite{folkes99}; \cite{slonim00}) used global analyses of the spectra
related to principal component analysis to define types.

The differences in type definitions, magnitude systems, and survey
volumes make this comparison difficult, but comparison of all the surveys
suggests a qualitative difference in the shapes of the luminosity functions 
derived from  morphological and spectral classification.  
Luminosity functions are usually parameterized by the
Schechter (1976\markcite{schechter76}) form, $dn/dL =
(n_*/L_*)(L/L_*)^\alpha \exp(-L/L_*)$ with a comoving density $n_*$,
break luminosity $L_*$, and a faint-end slope $\alpha$.  
The differences in luminosity functions derived by morphological and
spectroscopic classification are shown in Figure~1.
The surveys based on spectroscopic classification show less agreement
among them than those based on morphological classification.
The surveys using spectral typing also tend to find shallower
faint-end slopes for early-type galaxies and steeper slopes for
late-type galaxies---as well as larger differences between the 
two slopes---than do the surveys using morphological typing. 
These differences for normal galaxies must arise from one of two 
reasons:  either the differences point to intrinsic differences between
the methods and hence intrinsic differences among galaxy properties
(morphological versus spectroscopic), 
or one of the methods is systematically biased.

\begin{figure}[t]
\figurenum{1}
\epsscale{0.5}
\plottwo{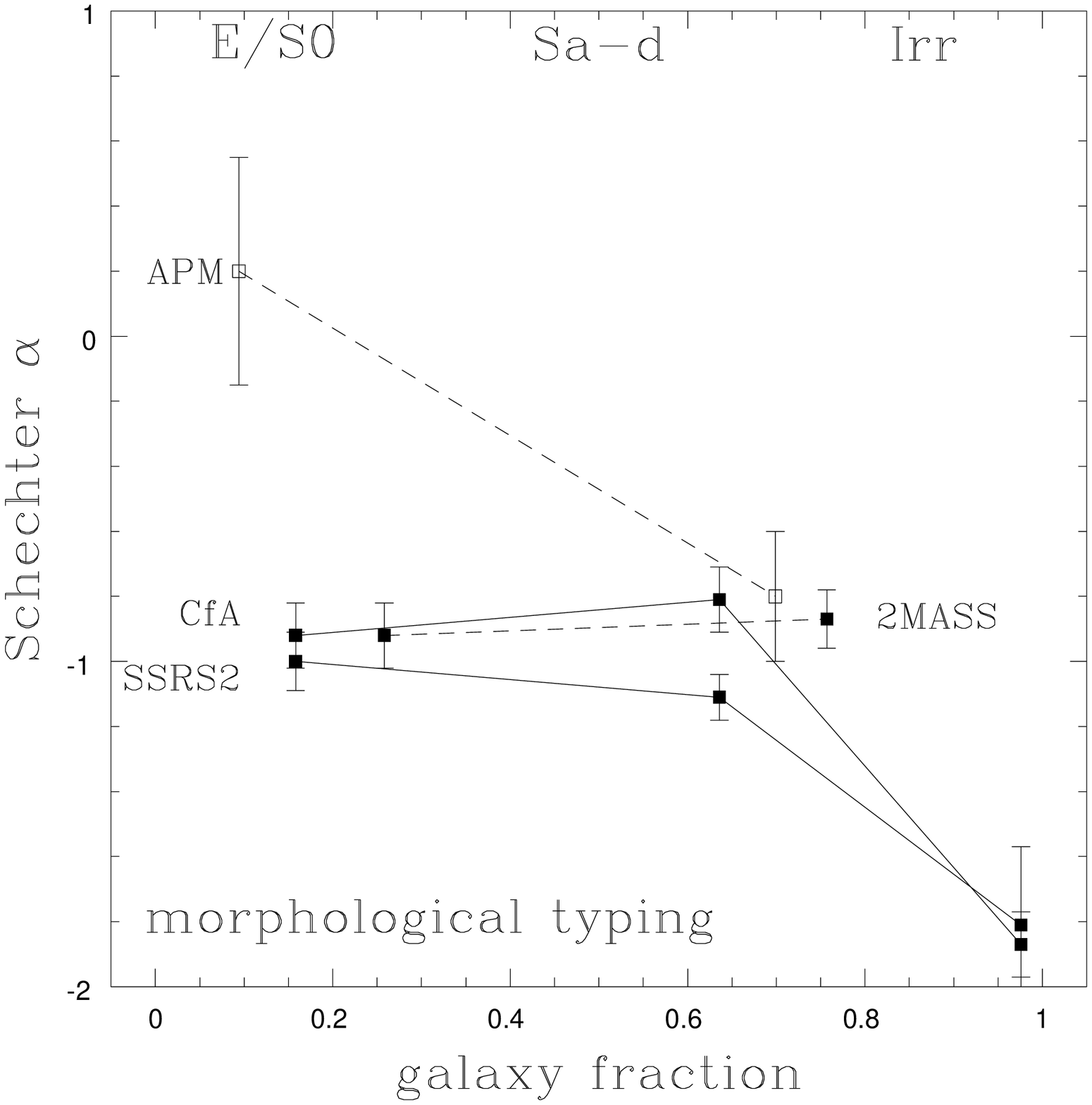}{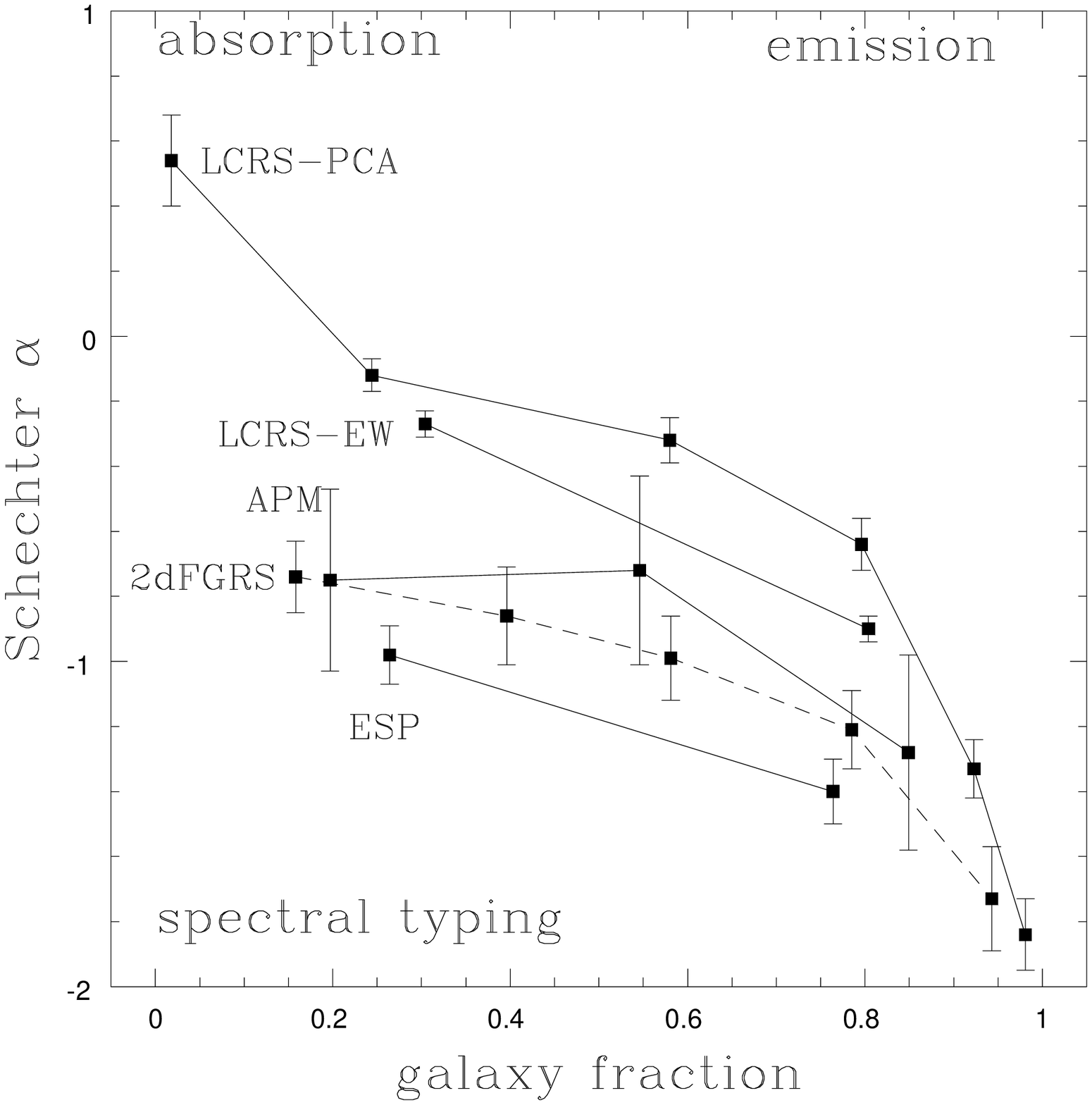}
\caption{
Luminosity function shapes for morphologically [left] and spectrally
[right] classified galaxy samples.  The Schechter $\alpha$
parameter (1976) is shown for each survey as a function of galaxy type
ordered from early to late type in going from left to right.  If
fraction $f_i$ of the galaxies in the survey are of type $i$, then the
galaxy fraction assigned to the type is $x_i = \sum_{j<i} f_j + f_i/2$
which is the sum of the number of galaxies in the earlier bins plus
half the number in the current bin.  The APM survey classified by
morphological type has 21\% of  its galaxies unclassified (which have
been added as a gap between the early  and late type results), and we
used the type fractions of the SSRS2 survey for  the very similar CfA
survey since Marzke \etal\ (1994b) 
did not report the numbers of galaxies in the survey for each type.
The studies based on morphological classification show better
consistency in their derived luminosity functions than studies based
on spectral classification.
\label{fig-lf-morph-spec}
  }
\end{figure}

\placefigure{fig-lf-morph-spec}

Suspicion about galaxy classification usually focuses on morphological classification because of its 
dependence on biological neural networks and its association with old-fashioned, uncomputerized
astronomy.  While many astronomers may feel that the definitions of morphological
galaxy types are similar to the famous, if nebulous, Supreme Court definition
of pornography (``I know it when I see it,'' \cite{stewart64}), 
the studies by Naim \etal\ (1995a\markcite{naim95a},b\markcite{naim95b})
demonstrated that the results are consistent between different implementations of biological neural networks and 
can be automated and reproduced using silicon neural networks.  Successful morphological classification
does require high dynamic range and resolution images which largely confines its use to the
local universe (with ground-based imaging) or intermediate redshifts (with the \emph{Hubble Space
Telescope}).  Morphological classification remains labor intensive and difficult to apply 
on a very large scale, which has driven most recent
surveys to switch to spectral classification.  Although the methods differ in their details, the basic 
assumption is that the \emph{global} spectra of galaxies form a continuous sequence between spectra 
dominated by absorption and emission lines and that this sequence is related to, but not necessarily identical to,
the classical morphological sequences (see \cite{connolly95}; \cite{zaritsky95}; \cite{bromley98a}; 
\cite{folkes99}; \cite{slonim00}).
The primary advantages of spectral typing are that it
is easily automated and that the spectroscopic data required for it are obtained in the 
course of the redshift survey.

Since, as Figure 1 illustrates for the faint-end slope, the parameters of the various luminosity
functions are mutually inconsistent, we started searching for tests and cross calibrations which
could be used to reconcile or check the results.  In Kochanek \etal\ (2000a\markcite{kochanek00a}) 
we used the 2MASS survey to compare directly the magnitude systems in the different surveys.
Here we apply a simple test for the self-consistency of luminosity functions by type---namely that the 
type fractions should be nearly independent of the apparent magnitude---to check whether systematic
errors in defining galaxy types can be responsible for the differences in the shapes of luminosity
functions.  
This test is described and applied in \S2 to the samples for which we could obtain the necessary data. 
The test is passed by the local, morphologically typed samples and failed by the spectrally typed
samples based on small aperture fiber optic spectra.  
The origin of the problem is described in \S3 as a probable consequence of the small spectroscopic apertures.  
The future of using spectra to assign galaxy types is summarized in \S4.

\section{A Simple Test for Problems In Galaxy Typing}

At low redshifts the galaxy populations can be characterized by a set of non-evolving 
differential luminosity functions $\phi_i(L)$ for each galaxy type $i=1\cdots N$.  It 
is easy to show that the type fraction 
is independent of apparent magnitude for galaxies with arbitrary luminosity functions 
(with identical $k$-corrections) in a homogeneous Euclidean universe.  
In real low redshift galaxy samples the deviations of the
real cosmological distances from Euclidean distances, differential $k$-corrections,
and type-dependent variations in the mean galaxy density with redshift all modify
this simple result, but the general rule that the galaxy type fractions should 
depend weakly on apparent magnitude still holds.  The differences between the
observed and predicted type fractions as a function of magnitude are a simple,
but powerful test for systematic errors in galaxy classification schemes. 

For a differential luminosity function $\phi_i(M)$ defined as a function of 
absolute magnitude $M = m - 5 \log \left(D_L(z)/r_0\right) - k_i(z)$
for apparent magnitude $m$, luminosity distance $D_L(z)$, $r_0 = 10$~pc, 
and type-dependent $k$-correction $k_i(z)$, 
the predicted type-dependent number counts are
\begin{equation}
  { d n_i \over d m } = \Delta\Omega \int_0^\infty \phi_i(M) D^2 dD 
\end{equation}
for a flat universe with a comoving volume element of $\Delta\Omega
D^2 dD$.  Type-dependent comoving density fluctuations modify the
integral by $\phi_i(M)\longrightarrow\delta_i(z)\phi_i(M)$, where
$\delta_i(z)$ is the fractional variation in the density of galaxy
type $i$.  While the density variations can significantly affect the overall
number counts, the type fractions,
\begin{equation}
        f_i(m) = { dn_i \over dm } \left[ \sum_{j=1}^N { dn_j \over dm
        } \right]^{-1},
\end{equation}
depend only weakly on density variations given the strength of the
morphology-density relation and the survey volumes.  The type
fractions averaged over modest magnitude windows ($\Delta m = 0.5$~mag)
include too much volume to be sensitive to these density effects.  We
test for inconsistencies by comparing the predicted and observed type
fractions as a function of apparent magnitude.  The test will \emph{underestimate} 
the significance of any differences because the
luminosity functions were derived from the same data.  For each survey
we calculated $dn_i/dm$ for each galaxy type and using the magnitude
definitions appropriate to the survey including the varying
type-dependent $k$-corrections.

Four recent redshift surveys have used morphological classification to
define luminosity functions by type.  In Figure 2 we compare the
observed and predicted morphological type fractions as a function of
apparent magnitude.  The APM survey (\cite{loveday92}, 1996\markcite{loveday96}) was
too deep compared to the quality of the photographic plates used for the
photometry to permit accurate morphological classification of the full
sample.   Indeed, 348 of the 1658 sample galaxies were not included in
computing the luminosity functions  by morphological type because they
were too compact or faint to estimate a morphological type.   As we
see in Figure 2, the type fractions depend strongly on magnitude and
are inconsistent with the predictions of the fitted luminosity
functions.  The generally accepted interpretation 
(see \cite{loveday92}, 1999\markcite{loveday99}; \cite{marzke94b}) 
is that the 348 galaxies without
classifications are preferentially  faint early-type galaxies so that the
observed drop in the early-type fraction for $b_J>16$~mag  is an
artifact.  The CfA (\cite{marzke94b}), SSRS2 (\cite{marzke98})
and 2MASS (\cite{kochanek00a}) luminosity functions are  broadly
consistent with their observed type fractions, albeit with local
anomalies, such as the Virgo cluster region, 
which are significantly larger than the statistical
uncertainties.  The two blue-selected surveys (CfA and SSRS2) show
very similar overall type fractions---although the spiral fraction is
20\% higher in the CfA survey, the difference is consistent with the
uncertainties in the density normalizations.  The infrared-selected
2MASS survey looks like it has very different type fractions, but this
is just a consequence of the change in the selection wavelength.

\begin{figure}[p]
\figurenum{2}
\centerline{\psfig{figure=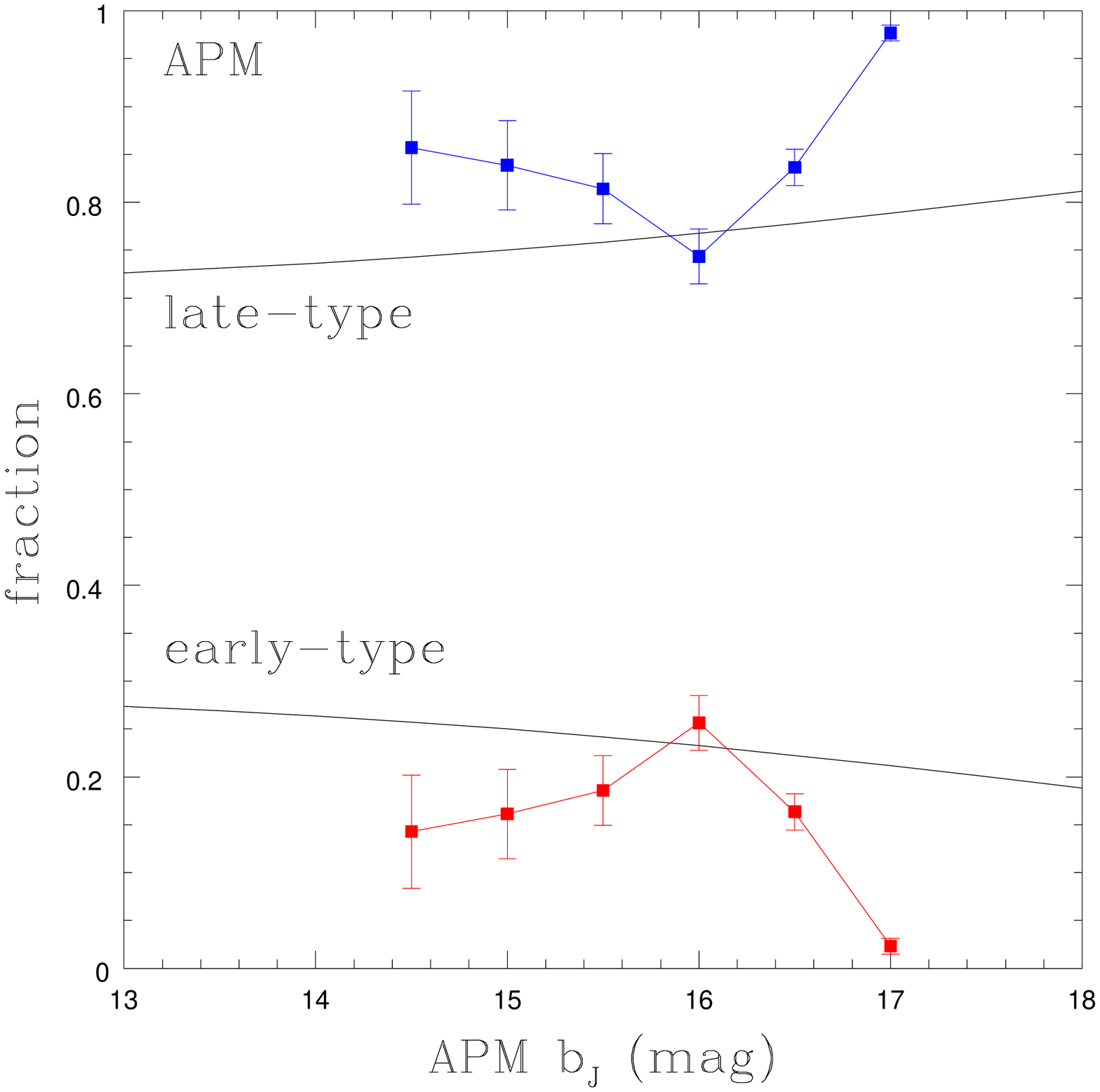,height=3.25in}\psfig{figure=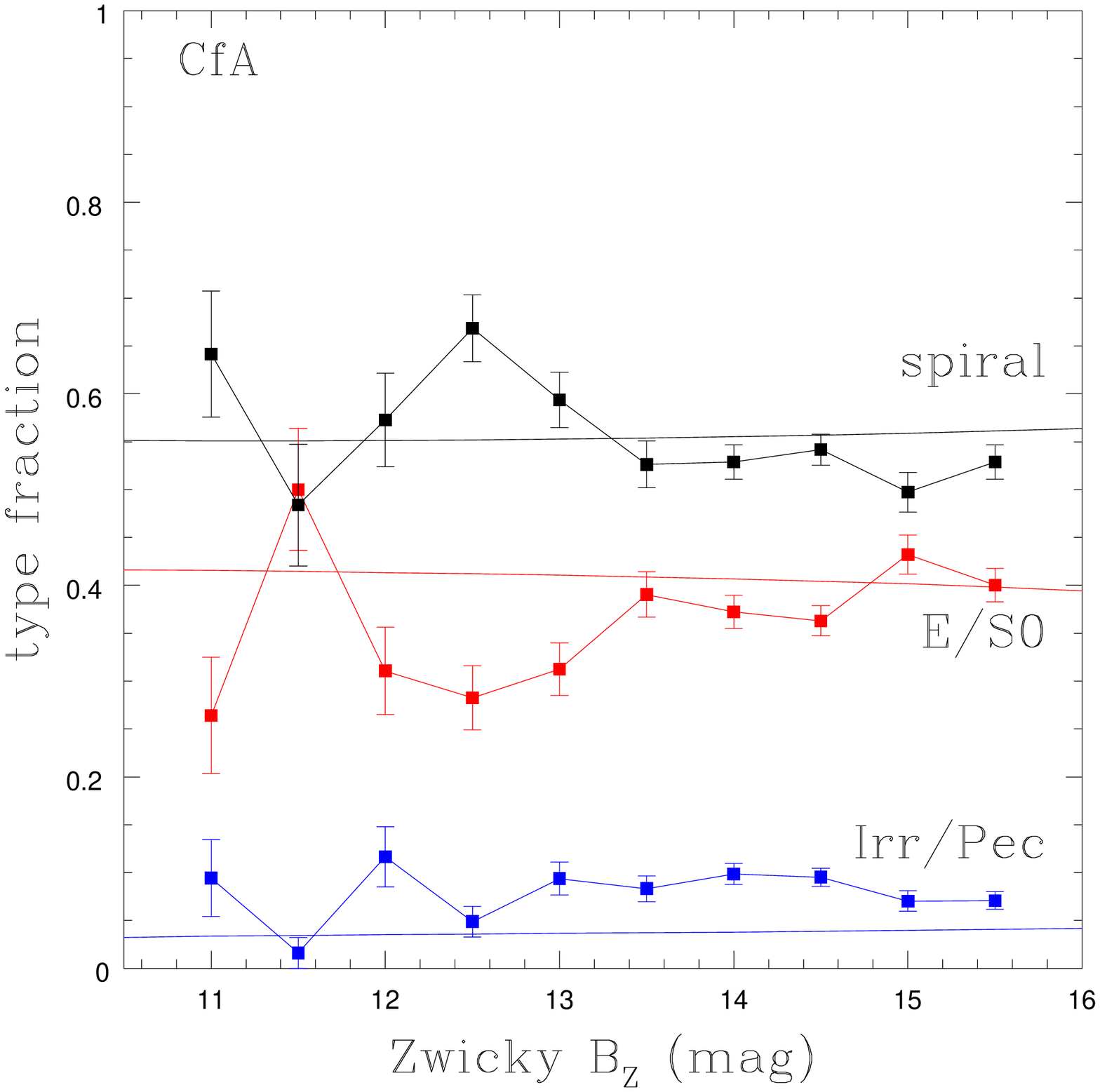,height=3.25in}}
\centerline{\psfig{figure=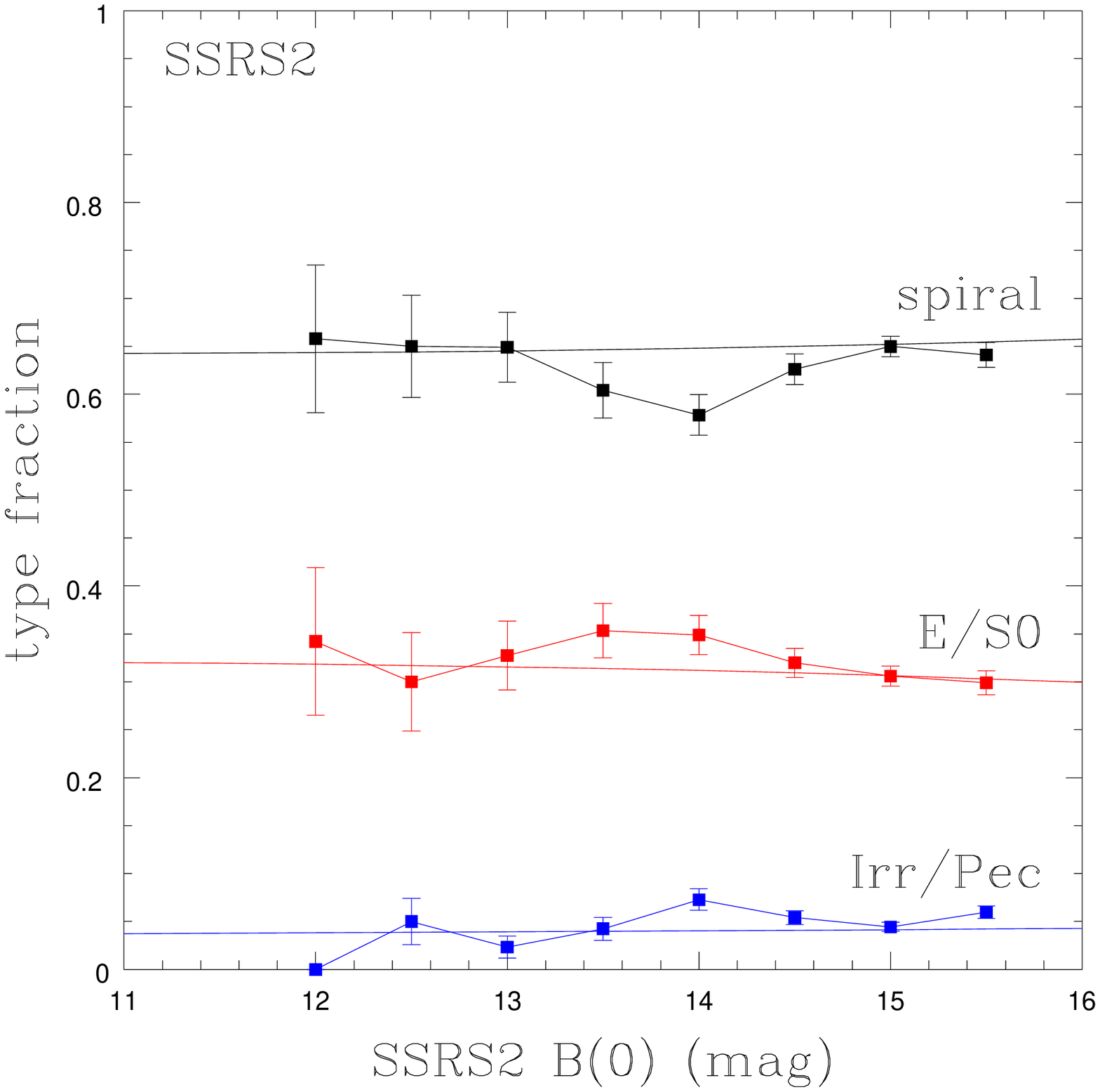,height=3.25in}\psfig{figure=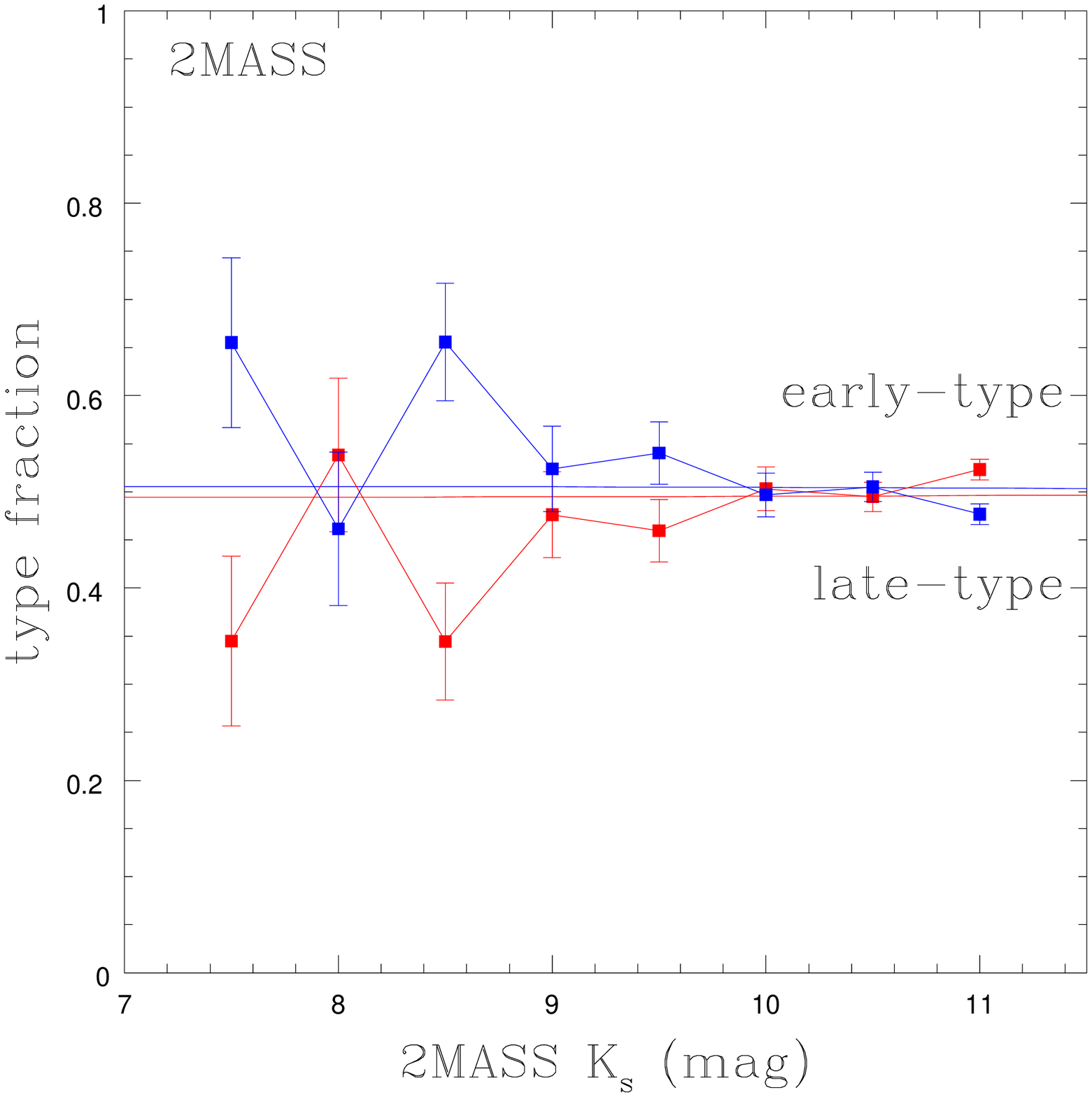,height=3.25in}}
\caption{Morphological type fractions as a function of apparent magnitude for the APM (top left), 
CfA (top right), SSRS2 (lower left) and 2MASS (lower right) surveys.  The points with error
bars are the data and the curves are the type fractions predicted by the best fit Schechter
luminosity functions for the surveys. For the APM survey the relative comoving densities
were set to produce the correct average type ratio, as Loveday \etal\ (1992) did not report
the density normalizations for their LFs by type.  For the CfA survey, the data we used
does not exactly match that fit by Marzke \etal\ (1994b) in deriving the luminosity 
function which may lead to differences in the mean comoving densities between the model 
and the data used here.  
The CfA and 2MASS survey regions overlap but were independently typed.  
Note that the features in the type fraction seen for bright magnitudes (which are dominated 
by the Virgo cluster region) are almost identical.
The CfA, 2MASS, and SSRS2 surveys all exhibit morphological type fractions which are,
as predicted, roughly constant with apparent magnitude.
\label{fig-morph-mag}
  } 
\end{figure}

\placefigure{fig-morph-mag}

We can make a similar comparison for three recent redshift surveys which defined spectral
types for the sample galaxies, as shown in Figure 3.  Tresse \etal\ (1999\markcite{tresse99}) 
and Loveday \etal\ (1999\markcite{loveday99})
reanalyzed the APM survey by dividing the sample into bins of low, medium, and high H$\alpha$ (and O[II]) 
equivalent widths.  The spectra were obtained with an 8~arcsec wide long-slit spectrograph which 
typically covered 45\% of the galaxy above the 25~mag/arcsec$^2$ isophote (see \cite{tresse99}). 
The LCRS (\cite{shectman96}) and ESP (\cite{vettolani97}) surveys used fiber spectrographs with 
fiber diameters of 3\farcs5 and 2\farcs5 respectively. 
Both the LCRS (\cite{lin96}) and ESP (\cite{zucca97}) luminosity functions
defined by emission line equivalent widths show strong systematic 
trends in the type fraction with redshift---the low EW fraction diminishes with redshift while the 
high EW fraction steadily rises.  In the Bromley \etal\ (1998a\markcite{bromley98a}) analysis of the LCRS survey, 
the galaxies were divided into 6 spectral clans
(1=earliest, 6=latest) which we have simplified in Figure 3 by combining the clans in pairs
since each pair member shows very similar behavior.  The earliest pair (clans 1 and 2)
shows a steep drop with magnitude, the intermediate pair (clans 3 and 4) shows a compensating
rise, and the latest pair (clans 5 and 6) shows no variation.  Except for the APM results 
and the latest clan pair in Bromley \etal\ (1998a\markcite{bromley98a}), the variations in the type fractions 
are in gross disagreement with those predicted from the luminosity function models.

\begin{figure}[p]
\figurenum{3}
\centerline{\psfig{figure=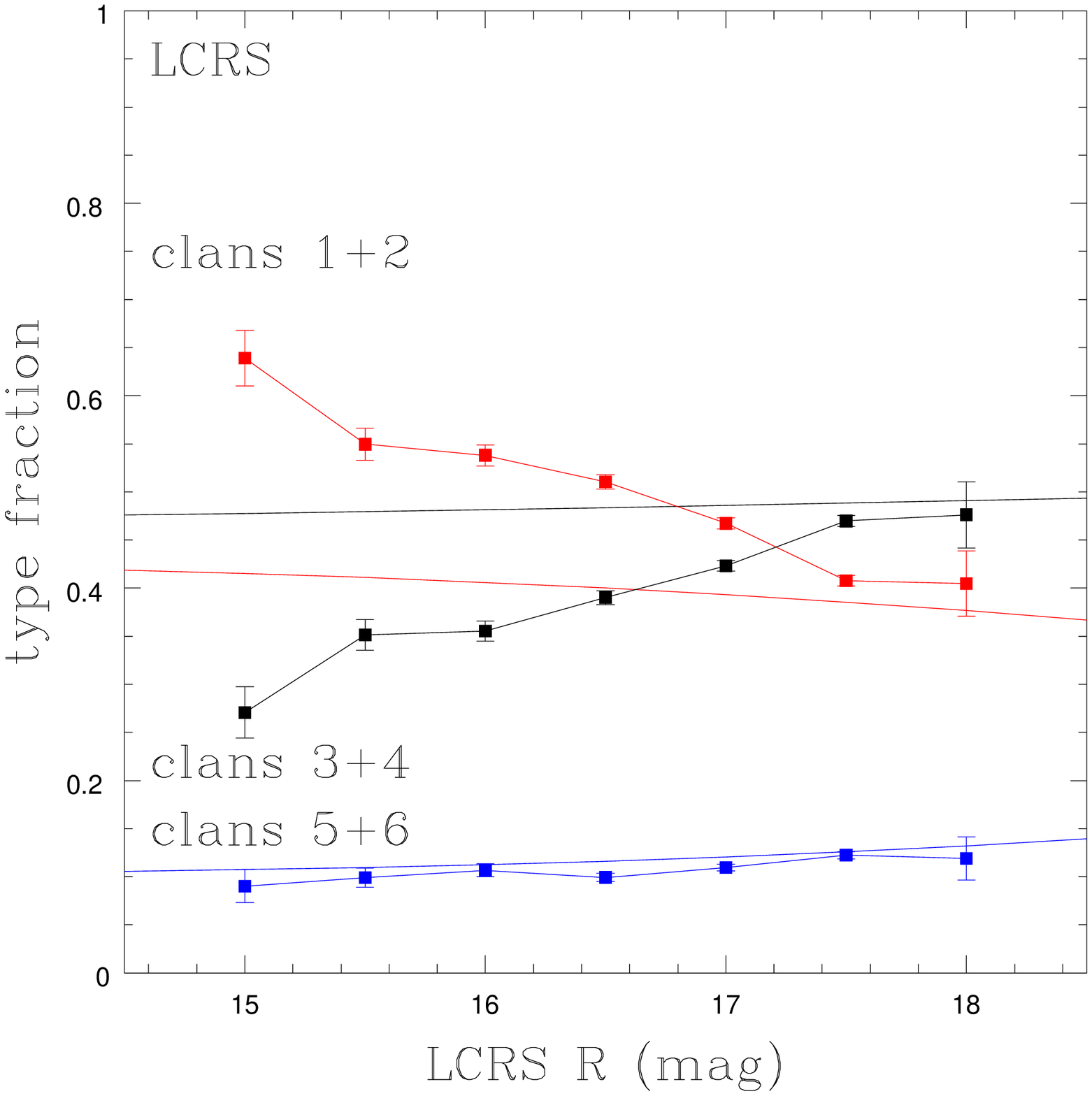,height=3.25in}\psfig{figure=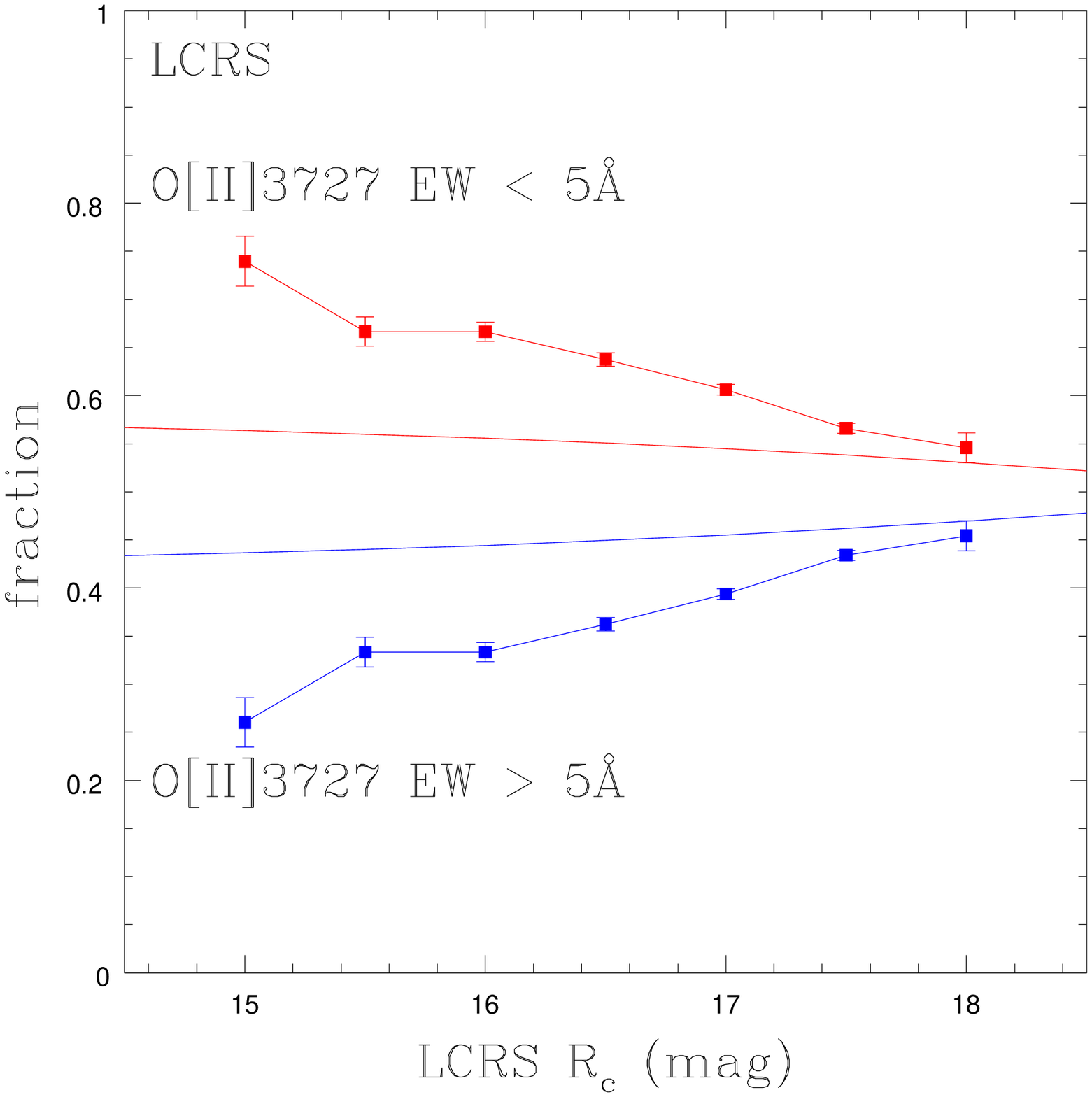,height=3.25in}}
\centerline{\psfig{figure=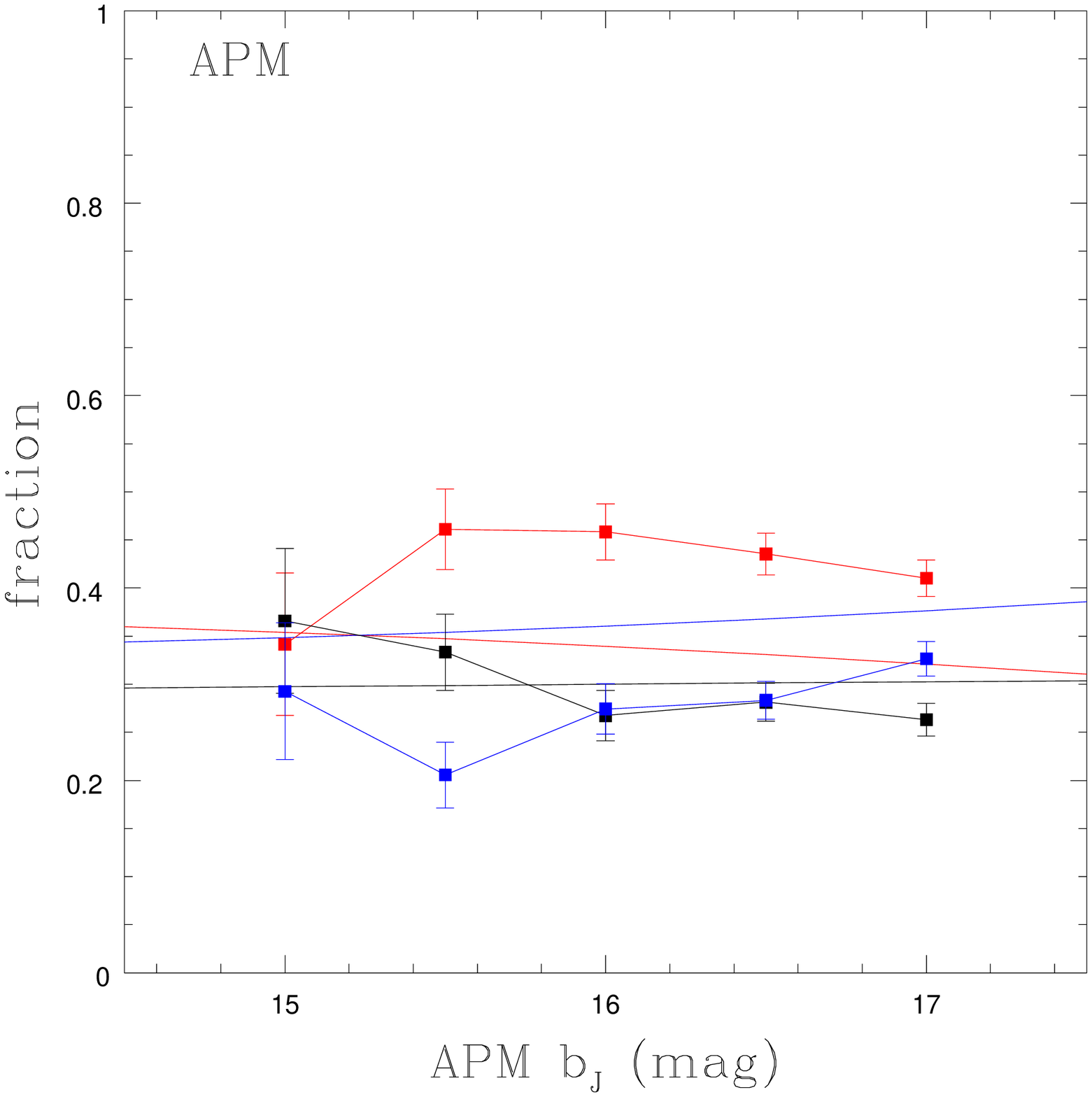,height=3.25in}\psfig{figure=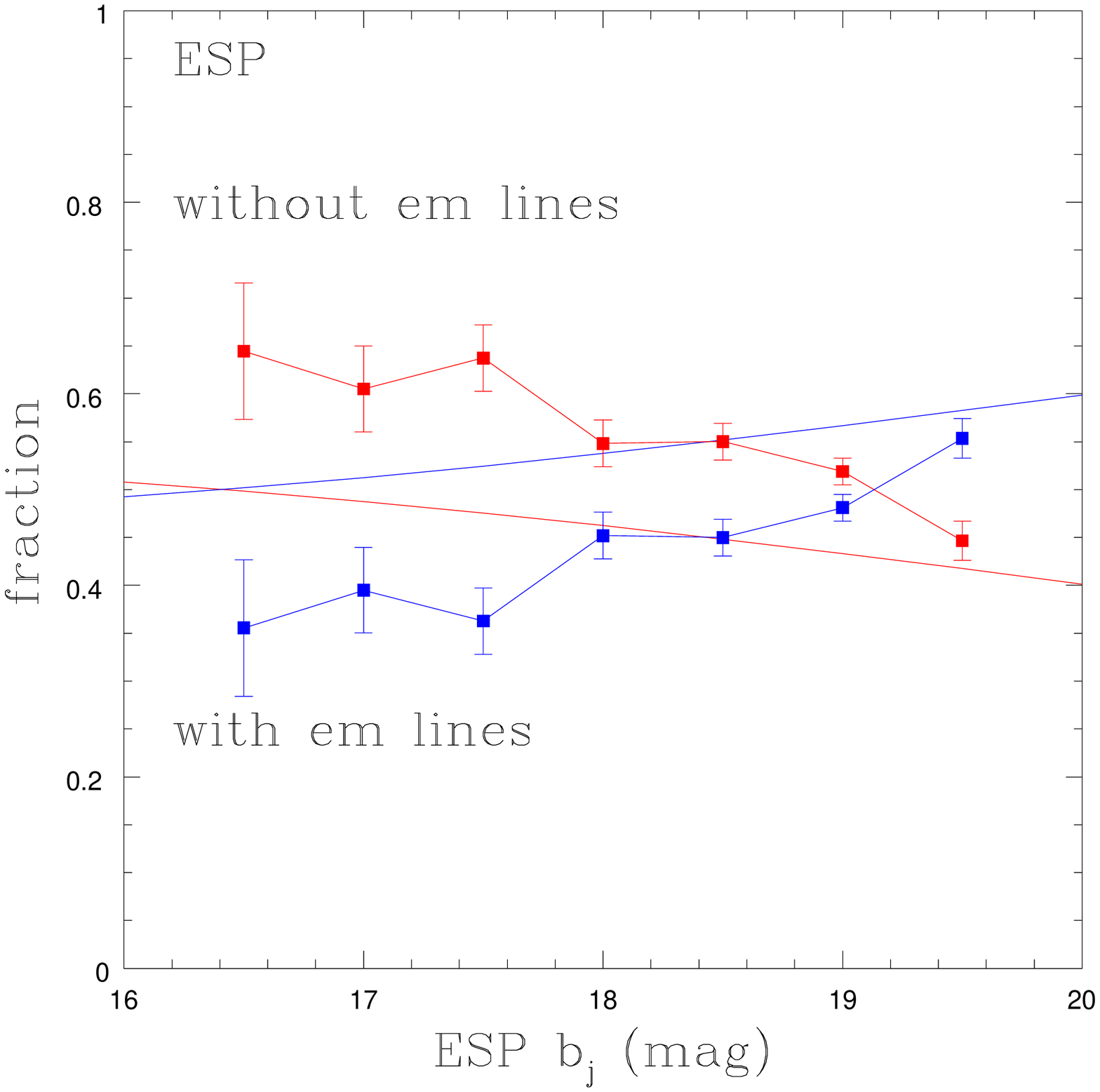,height=3.25in}}
\caption{Spectral type fractions as a function of apparent magnitude for the LCRS by
spectral clan (top left) and emission line equivalent width (top right), for the 
APM survey by emission line equivalent width (lower left), and the ESP survey by
emission line equivalent width (lower right). 
The LCRS (classified both by ``clans'' and [O~II] equivalent widths) and ESP surveys,
which were all done with fiber-fed spectrographs,
show strong dependences of spectroscopic type fractions with apparent magnitude;
the APM survey, which was based on large aperture long-slit spectra, does not.
\label{fig-spec-mag}
}
\end{figure}

\placefigure{fig-spec-mag}

\section{The Failure of Spectral Typing}

While no survey has type fractions which exactly match the predictions
of its luminosity functions, we conclude that morphological
classification leads to broadly self-consistent results for bright,
nearby  galaxy samples (CfA, SSRS2, 2MASS). It can fail badly for more
distant, fainter samples like the APM when the images lack sufficient
dynamic range or resolution to distinguish successfully among galaxy types.
The surprising result is that spectrally-typed samples from fiber
surveys (ESP and LCRS)  fail the test badly, while the smaller APM
sample based on long slit spectra passes the test.  Global spectra of
galaxies must lead to well-defined, self-consistent spectral type
definitions, so  the most likely origin of the problem is the small
size of the spectral apertures used by the fiber Surveys.  At the
median redshift ($z\simeq 0.1$) of all the fiber surveys (LCRS, ESP,
2dFGRS, SDSS)  the fiber sizes match the typical size of a galactic
bulge ($1\farcs0=1.2 h^{-1}$~kpc). 
In the fiber surveys, an early-type spiral galaxy with a large bulge will have the spectrum of an
early-type galaxy when it is bright, nearby, and the fiber covers only
the bulge; on the other hand, it will have a later-type  spectrum when it is faint,
distant, and the fiber includes part of the disk.
Such an \emph{aperture bias} very naturally explains the trends seen for
the ESP and LCRS surveys in Figure 3.

Few studies using spectral typing methods have considered the
problems of aperture bias.  Tresse \etal\ (1999\markcite{tresse99})
discuss the issue for the APM survey, concluding that their 8\farcs0
wide longslit aperture is large enough to avoid any biases, as is
borne out by our tests in \S2.  Zaritsky  \etal\
(1995\markcite{zaritsky95}) discuss the issue briefly, concluding that
aperture bias would not be a problem.  Our own reading of the Zaritsky
\etal\ (1995\markcite{zaritsky95})  tests drives us to the opposite
conclusion, at least for  the restricted problem of determining
luminosity function shapes where differential errors in
classifications  as a function of luminosity can radically change the
derived faint-end slopes.  Quantitatively estimating the extent of the
problem is difficult because we lack detailed probability
distributions for the structural variables describing spiral galaxies
(bulge-to-disk luminosity ratios, disk scale lengths, and bulge scale
lengths as a function of total luminosity).  We can, however, show
that the spectral types are ill-defined by comparing the fiber
apertures to images of galaxies in \S3.1, examining the variation in
H$\alpha$ equivalent widths with aperture size in \S3.2, and
considering simplified models of galaxies in \S3.3.  In \S3.4 we
explore some of the consequences of aperture bias for spectrally-typed
luminosity functions.

\subsection{Evidence for Aperture Bias from Galaxy Images}

We start by illustrating the problem with images.  Figure 4 shows the
3\farcs5 diameter LCRS fiber aperture  superposed on
POSS~II\footnote{The Second Palomar Observatory Sky Survey (POSS-II)
was made by the California  Institute of Technology with funds from
the National Science Foundation, the National Geographic Society, the
Sloan Foundation, the Samuel Oschin Foundation, and the Eastman Kodak
Corporation. The Oschin Schmidt  Telescope is operated by the
California Institute of Technology and Palomar Observatory.  }
\footnote{The Digitized Sky Surveys were produced at the Space
Telescope Science Institute under U.S. Government grant NAG
W-2166. The images of these surveys are based on photographic data
obtained using the Oschin Schmidt Telescope on  Palomar Mountain and
the UK Schmidt Telescope. The plates were processed into the present
compressed  digital form with the permission of these institutions. }
images of Sa and Sb galaxies with a range of absolute luminosities
($-2$~mag to $+2$~mag relative to $L_*$ from our 2MASS survey) after
rescaling the images to  the redshift at which they would have
$R_c\simeq 16.5$~mag in the LCRS survey.  The images make clear that
the fiber apertures sample only central spectra and that the region
covered by the aperture depends strongly on redshift or apparent
magnitude.  Even for the faintest LCRS galaxies ($R_c \simeq 18$~mag),
most of the galaxy disks lie outside the aperture.  The aperture
problem becomes steadily worse for the SDSS, ESP, and 2dFGRS surveys
which have smaller fiber diameters (3\farcs0, 2\farcs5,
and 2\farcs0 respectively) but approximately the same median redshift.
The biases become smaller for surveys with larger spectrographic
apertures, as illustrated by the APM results.   Seeing (which will
increase the effective fiber aperture by 0\farcs5 to 1\farcs0
depending on the site and the observing conditions) and
pointing/coordinate errors ($\sim 0\farcs3$ to $1\farcs0$) will
increase  the disk fraction entering the fiber, but not by enough to
remove the problem. For example, we left the
pointing offsets between POSS~II and 2MASS in Fig. 4 to provide an
indication of their importance.
Non-uniformities in the fiber illumination will
tend to reduce the effective aperture.  

\begin{figure}[p]
\figurenum{4}
\epsscale{0.9}
\plotone{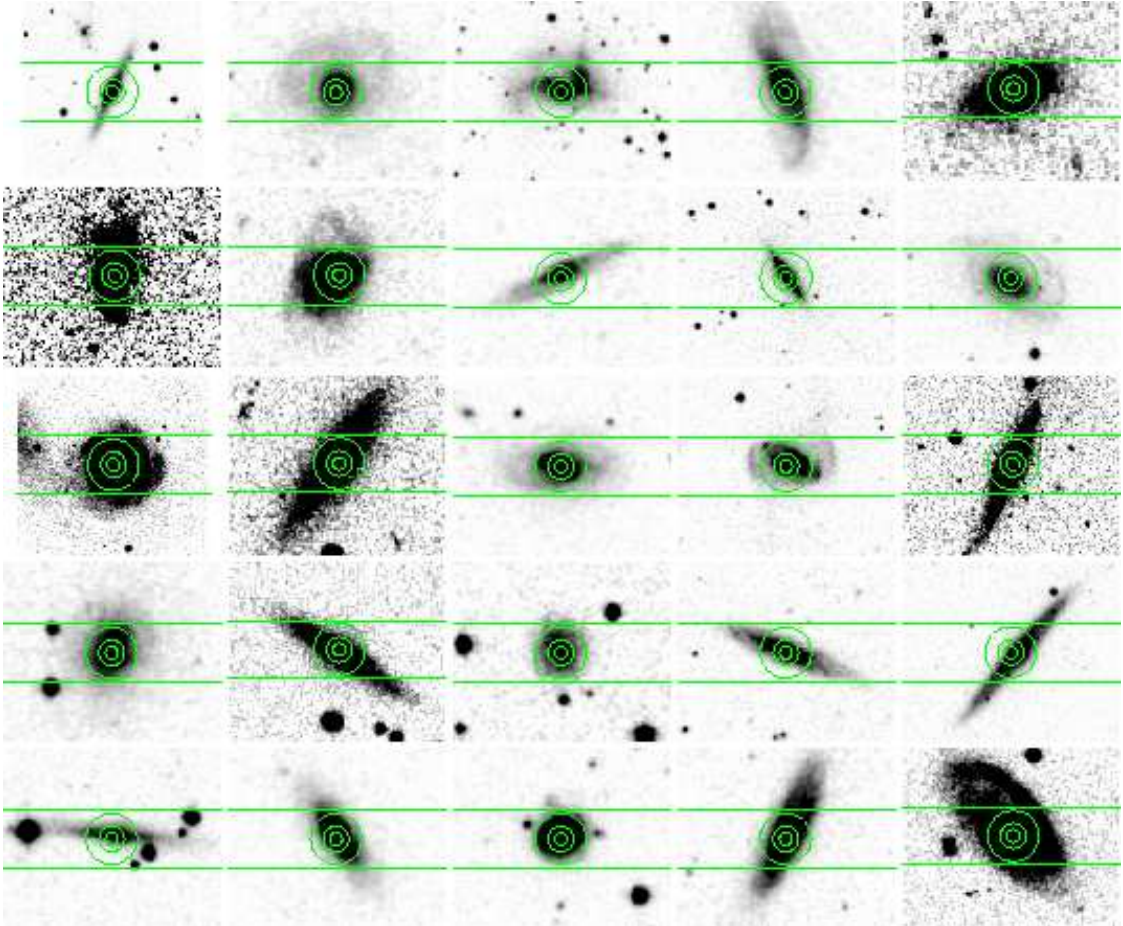}
\caption{
Fiber apertures superposed on galaxies.  The greyscale POSS~II images
are Sa and Sb galaxies  from the 2MASS survey rescaled to the
redshifts they would have at the midpoint of the LCRS survey
($R_c=16.5$~mag).  The galaxies in each row have the same absolute
luminosity ($-2$, $-1$, $0$, $+1$ and $+2$~mag relative to $L_*$ from
top to bottom) based on the late-type galaxy infrared luminosity
function.  The superposed circles show the area covered by the
3\farcs5 diameter LCRS fiber aperture at the redshifts corresponding
to the galaxy being observed over the magnitude range of the LCRS
survey.  The offsets are due to astrometry differences between 2MASS
and the digitized POSS~II images, which we left in to give a feeling
for the effects of pointing errors.  The largest circle corresponds to
the faint limit, $R_c=18$~mag,  the middle circle to the midpoint
$R_c=16.5$~mag, and the smallest circle to the bright limit
$R_c=15$~mag.  The ESP, 2dFGRS and SDSS surveys have very similar
survey depths but smaller fiber aperture  diameters of 2\farcs5,
2\farcs0 and 3\farcs0 respectively.  The horizontal lines show the
geometry of the APM spectrograph slit at the LCRS midpoint and should
be compared in size to the intermediate circle.  It is clear that the
typical fiber size (intermediate circles) represent incomplete and
unrepresentative samples of the light distribution of the galaxies; it
would be extremely difficult to classify these galaxies accurately if
only the light within the intermediate circles were used.  On the
other hand, the large rectangular apertures of the APM survey are a
far better sampling of the overall galaxy light distributions, and
hence classifications based on such spectra are arguably more
reliable, accurate, and representative.
\label{fig-image-apertures}
    }
\end{figure}

\placefigure{fig-image-apertures}

\subsection{Evidence for Aperture Bias from Galaxy Spectra}

We can also illustrate the problem using spectra.  We have a large
sample of high signal-to-noise ratio, major axis, long slit spectra of
2MASS galaxies in the magnitude range $10.5 < K_s < 11.25$~mag, which were
obtained to measure the dynamical properties of a statistically
well-defined sample of galaxies.  The slit width is 2\farcs0. We
extracted spectra for slit lengths of 2\farcs6, 5\farcs2, 10\farcs3,
20\farcs6 and 41\farcs3  (2, 4, 8, 16 and 32 pixels) and computed the
H$\alpha$  equivalent width for each aperture.  The median redshift of
galaxies in this magnitude range is $z\simeq 0.027$
(\cite{kochanek00a}) which roughly corresponds to $13.4 < R_c
<14.2$~mag in the LCRS sample if we match the two samples using median
redshifts.   
Galaxies were classified as having central AGN if their emission-line equivalent
widths were $\log$[O~I]$/\log$H$\alpha > -1.2$~dex or $\log$[N~II]$/\log$H$\alpha > -0.25$~dex, 
which we adapted from Osterbrock (1989\markcite{osterbrock}) to match those lines available
in our spectra.
``Star-forming'' galaxies were defined as those with detectable H$\alpha$ in all apertures;
``intermediate'' galaxies were those without H$\alpha$ in one or more apertures.
Figure 5 shows that many galaxies with emission lines
have equivalent widths which rise with increasing aperture area,
although there is also a minority of star-forming galaxies with falling equivalent
widths which could be created by emission from a weak central AGN.    For
illustration we can count the number of galaxies crossing the spectral
boundaries used by the APM survey at H$\alpha$ equivalent widths of
2\AA$\,$ and 15\AA.  Of the 76 galaxies we studied here, seven (9\%)
cross from low  to intermediate EW, five (7\%) cross from intermediate
to high EW, seven (9\%) cross from  intermediate to low EW, and three
(4\%) cross from high to intermediate EW.   In total, 29\% of the
galaxies change their spectral classification as we increase the slit
length.
Note that the slopes of many galaxies with decreasing EW become fairly flat
for the longer slit lengths, while the slopes of galaxies with increasing EW are
often still increasing rapidly for the longer slit lengths.

\begin{figure}[p]
\figurenum{5}
\epsscale{0.8}
\plotone{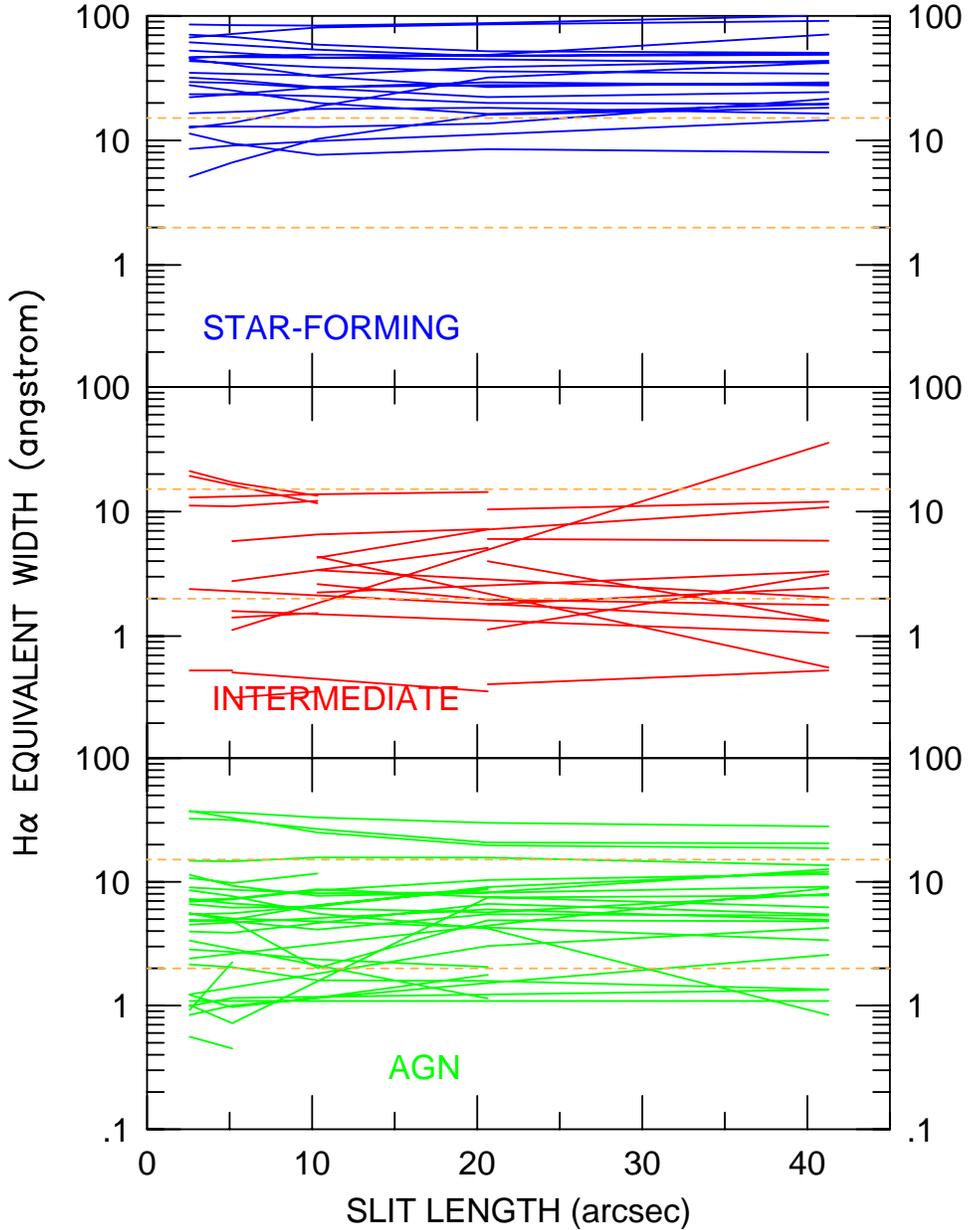}
\caption{
The H$\alpha$ equivalent widths as a function of spectroscopic aperture length
for 76 randomly-selected 2MASS galaxies with $10.50 < K_s < 11.25$~mag.
The dashed horizontal lines are the spectral type boundaries used by the APM survey.
Galaxies were roughly subdivided by spectroscopic type:  ``AGN'' are those galaxies with
high [O~I]/H$\alpha$ or [N~II]/H$\alpha$ in their central aperture; 
``star-forming'' galaxies were those with detectable H$\alpha$ in all extracted apertures;
 and ``intermediate'' are those with H$\alpha$ not detected in one or more aperture.
A significant percentage of galaxies, with and without AGN, cross the APM type boundaries
with increasing spectroscopic aperture, thus demonstrating a potential aperture bias
in spectroscopic classification.
\label{fig-mf-spectra}
    }
\end{figure}

\placefigure{fig-mf-spectra}

Geometry makes it difficult to convert quantitatively our observations
into equivalent LCRS  observations.  Let $r$ be the ratio of the slit
length to the fiber diameter.  The spectrum of a 2MASS galaxy at
redshift $z_0$ covers the same area as the LCRS spectrum of a galaxy
at redshift $z=0.85 z_0 r^{1/2}$ and it has the same physical length
as the LCRS fiber  diameter at redshift $z=z_0 r$.  For a face-on
galaxy, matching the physical areas of the spectral apertures is the
better approximation (although the slit extends further into the disk
for the same total area), while for an edge-on galaxy, matching the
physical  lengths of the spectral apertures is the better
approximation.  Given that the median  redshift of the 2MASS sample is
only one-quarter than of the LCRS sample, $r\simeq 20$ corresponds to
the median LCRS redshift if we match the physical areas of the
apertures and $r\simeq 4$ if we match the physical lengths.  Thus,
Figure 5 either illustrates the equivalent width trends for the bright
LCRS sample, if you match physical areas, or the full LCRS sample, if
you match physical lengths.  

Figure 5 also illustrates the dangers of subdividing the sample into
large numbers of spectral types.  While only 29\% of the galaxies
would cross the APM boundaries  dividing the sample into three types,
a higher percentage would cross sample boundaries if we  divide the
sample into five types by adding another boundary at 5\AA.   Larger
numbers of classes lead to lower reliability because larger fractions of the
galaxies of a particular type are actually galaxies drifting through
that type due to the effects of aperture bias rather than galaxies
which intrinsically  have that spectral type.

\subsection{Evidence for Aperture Bias from Galaxy Models}

Finally, the problem can be illustrated with simple models.  We phrase the discussion in terms
of emission line equivalent widths, but the same effects will be found in more sophisticated
principal component analyses or galaxies typed based on broad band colors measured in small
apertures.  Consider a face-on galaxy composed of an exponential disk (scale length $h_d$)
and a de Vaucouleurs bulge (effective radius $R_e$) with a bulge-to-disk luminosity ratio of $B/D$. 
The average relation between the two scale
lengths is $\log R_e=0.95\log h_d-0.86$ ($\pm50\%$) for the K-band, but the scatter increases
steadily towards shorter wavelengths (\cite{dejong96}).  We assume that the spectrum of the disk
includes an emission line with equivalent width $EW_0$ while the bulge has no emission lines,
so that the global spectrum of the galaxy would have equivalent width $EW_g=EW_0/(1+B/D)$ 
because the addition of the continuum from the bulge reduces the equivalent width of the disk emission line.
In a finite radius spectral aperture $\theta$ ($1\farcs75$ for LCRS), we collect fractions
$f_D(\theta D_A(z)/h_d)$ and $f_B(\theta D_A(z)/R_e)$ of the light from the disk and the
bulge respectively, where $D_A(z)$ is the angular diameter distance to the galaxy.  We then
observe an emission line equivalent width of
\begin{equation}
       EW_{obs} =  EW_0 \left[ { 1         \over 1 + (f_B/f_D)(B/D) } \right]
                =  EW_g \left[ { 1 + B/D   \over 1 + (f_B/f_D)(B/D) } \right],
\end{equation}
which differs from the global value of the equivalent width.  Spectral types will be inconsistent
if the ratio $EW_{obs}/EW_g$ strongly on the intrinsic or extrinsic properties of the 
galaxy, and they will be biased if the ratio depends on luminosity.\footnote{The same
mathematics describes the effects in a principal component analysis (PCA).  Suppose the 
mean galaxy spectrum is $\vs_0$ and that the second component spectrum is $\vs_1$ such that 
a pure disk galaxy has spectrum $\vs_D=\vs_0+\vs_1$ and a pure bulge galaxy has a spectrum 
$\vs_B=\vs_0-\vs_1$.  A galaxy with a disk and a bulge has $D \vs_D+B\vs_B$, which we would
classify based on its normalized normalized $\vs_1$ spectral component, $t_{PCA}=(D-B)/(D+B)$,
where a pure disk has $t_{PCA}=1$ and a pure bulge is $t_{PCA}=-1$.  Through a fiber we observe 
spectrum $f_D D \vs_D+ f_B D \vs_B$, which has a PCA classification 
$t_{PCA}=(f_D D- f_B B)/(f_D D + f_B B)$ that is earlier than the true global spectrum. 
In short, the mathematics of aperture bias in PCA classifications is almost identical to that 
for the emission line equivalent widths.}  

As we qualitatively predicted based on Figures 4 and 5, the spectral properties of the galaxies as traced
by the variations in $EW_{obs}/EW_g$ depend strongly on the distance to the galaxy. Figure 6 shows the 
expected changes in $EW_{obs}/EW_g$ for galaxies with disk scale lengths of $h_d=3h^{-1}$~kpc and 
$0.5h^{-1}$~kpc and a range of bulge-to-disk ratios from 5\% to 30\% assuming the mean bulge scale 
length from de Jong (1996\markcite{dejong96}) and the LCRS fiber aperture.  We assumed that the $3h^{-1}$~kpc galaxy
was an $L_*$ galaxy with $M_R = -20.3$~mag and that the disk surface brightness was independent of
luminosity so that the $h_d=0.5h^{-1}$~kpc galaxy has $M_R=-16.4$~mag.  The larger the bulge-to-disk
ratio, the larger the fractional changes in the ratio over the redshift range for which the galaxy would
be included in the survey.  For a galaxy with $B/D=0.2$, a galaxy with an intrinsic equivalent
width of $EW_g=10$\AA\, would be classified as low equivalent width ($EW <5$\AA\, in \cite{lin96}) 
over half the survey's magnitude range, and the equivalent width is well below the true,
global value at all magnitudes.  This is, of course, a simplified example.  Some early-type
spirals show no emission lines along the line-of-sight through the bulge, which would exacerbate the effect.  
Edge-on galaxies would have a larger fraction of the disk light in the aperture if they 
were transparent but might not if we include dust.  Qualitatively, however, aperture bias
naturally produces the decreasing early-type and rising late-type galaxy fractions as a 
function of apparent magnitude in the LCRS and ESP surveys.

\subsection{Consequences of Aperture Bias for Luminosity Functions }

It is more difficult to determine whether the spectral types and the luminosity function 
are biased because of the sensitivity of the problem to so many parameters.  The simplest
mathematical description of the problem, neglecting inclination and extinction, starts 
from the distribution of galaxies in luminosity, spectral type (as represented by the
global equivalent width) and structural properties ($h_d$, $R_e$ and $B/D$), 
$ d n / dL dEW_g dh_d dR_e d(B/D)$.  We want to derive the distribution in luminosity
and spectral type
\begin{equation}
    { d n \over d L dEW_g } = \int dh_d dR_e d(B/D) { d n \over dL dEW_g dh_d dR_e d(B/D)},
\end{equation}
but we actually measure $EW_{obs}$ to obtain
\begin{equation}
  { d n \over d L dEW_{obs} } =  
       \int dh_d dR_e d(B/D) { d n \over dL dEW_g dh_d dR_e d(B/D)} { 1 + B/D \over 1 + (B/D)(f_B/f_D)}.
\end{equation}
where the correction factor is the ratio from eqn. (3) and illustrated in Fig. 6.  We can
simplify and illustrate the problem by assuming that the scale lengths are simple power laws
of the luminosity, $h=h_0(L/L_0)^\beta$ and $R_e=R_0(L/L_0)^\gamma$, so that
\begin{equation}
    { d n \over d L dEW_{obs} } = \int d(B/D) { d n \over dL dEW_g d(B/D)} { 1 + B/D \over 1 + (B/D)(f_B/f_D) }
\end{equation}
where the arguments for the curve of growth functions have the form $\theta D_A(z)/h_0 (L/L_0)^\beta$. 
In a Euclidean universe (which is almost correct locally), a source of flux $F$ and luminosity $L$ is
seen at distance $D_A(z)=(L/4\pi F)^{1/2}$, so the arguments for the curve of growth functions have
the form, $\theta (F/F_0)^{-1/2} (L/L_0)^{1/2-\beta}$ where $F_0=L_0/4\pi h_0^2$.  For \emph{self-similar}
galaxies, defined by $\beta=\gamma=1/2$, the arguments become independent of luminosity and depend
only on the flux.  This flux dependence is sufficient to explain the problems in current
spectral classifications.  Self-similar galaxies also have self-similar surface brightness profiles with
constant central surface brightnesses.    The de Jong (1996\markcite{dejong96}) mean K-band relation between
the disk and bulge scale lengths satisfies $\beta \simeq \gamma$ since $R_e \propto h_d^{0.95}$,
but the scatter is large and the relations for constant $EW_g$ and $B/D$ could differ from the
relation averaged over these variables.  The value of $\gamma$ is more problematic since it is
closely related to the problem of surface brightness completeness in surveys (e.g. \cite{dalcanton98}).  In standard
bright galaxy samples, all disks have similar central surface brightnesses (even if it is a
selection effect!) suggesting that $\beta\simeq 0.5$.  Unfortunately, the calculation is
extraordinarily sensitive to the value of $\beta$, as illustrated Figure 7 by using $\beta=0.4$,
$0.5$, and $0.6$ to scale the properties of the smaller galaxy from Figure 6.  In short,
aperture biases are almost certainly luminosity dependent, but we cannot prove (or disprove)
it from the available statistical data on the structural properties of spiral galaxies.

\begin{figure}[t]
\figurenum{6}
\epsscale{0.5}
\plotone{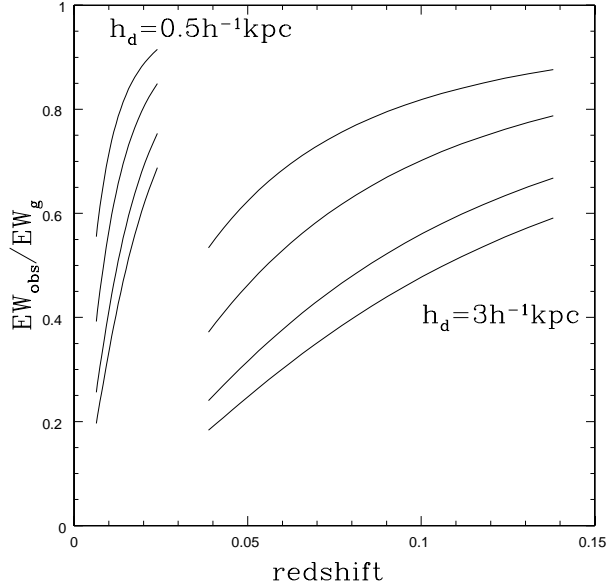}
\caption{The ratio of the measured to the global emission line equivalent width
  as a function of redshift for observations scaled to resemble the LCRS survey.  The
  bulge-to-disk ratios are 5\%, 10\%, 20\% and 30\% from top to bottom and the lines
  are plotted for the redshift range over which the galaxies would have $15 < R_c < 18$
  in the LCRS survey.
The aperture bias in measured equivalent width is a strong function of scale-length, 
bulge-to-disk ratio, and redshift.
\label{fig-eqwidth-redshift}
    }
\end{figure}

\placefigure{fig-eqwidth-redshift}

\begin{figure}[t]
\figurenum{7}
\epsscale{0.5}
\plotone{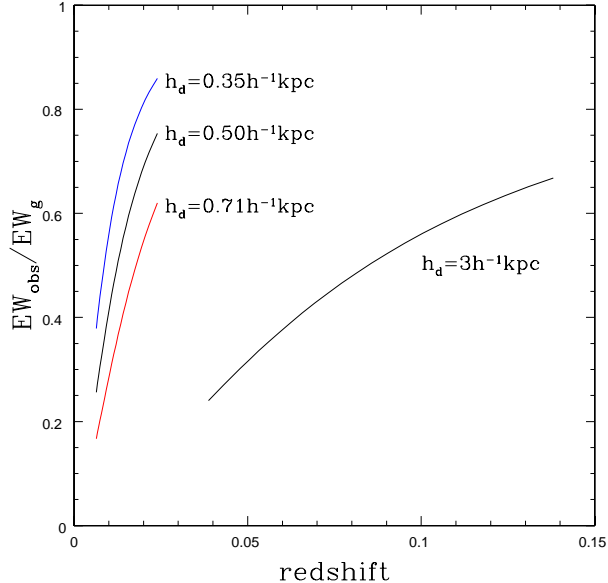}
\caption{Effect of deviations from self-similar surface brightness profiles.   We show the ratio
  of the observed and intrinsic equivalent widths for the $B/D=0.2$ case from Figure 6, but for
  the small galaxy we show the results of scaling the disk radius with luminosity as 
  $h\propto L^{\beta}$ with $\beta=0.6$ ($h_d=0.35 h^{-1}$~kpc), the self-similar exponent
  $\beta=0.5$ ($h_d=0.50 h^{-1}$~kpc), and $\beta=0.4$ ($h_d=0.71 h^{-1}$~kpc) corresponding
  to a change in the mean surface brightness by $\pm 0.75$~mag/arcsec$^2$ for a 3.9~mag
  change in the luminosity.
\label{fig-self-similar}
    }
\end{figure}

\placefigure{fig-self-similar}

The most likely effect of aperture bias is to misclassify fraction $f(M)$ of spectrally
late-type galaxies as early-type galaxies, and we can use this to explore the sensitivity
of spectrally-typed luminosity functions to its effects. The observed luminosity functions are 
\begin{equation}
   \phi_E^{obs}(M) = \phi_E(M) + f(M) \phi_L(M) \qquad\hbox{and}\qquad 
   \phi_L^{obs}(M) = \left[ 1-f(M) \right] \phi_L(M)
\end{equation}
rather than the true luminosity functions $\phi_E(M)$ and $\phi_L(M)$, but we can easily
invert the equation given a model for the error rate $f(M)$. Lets initially consider
the case where $f(M)=f$ is a constant.  The consequences of the errors then depend on the 
relative shapes of the luminosity functions.  If the shapes of the luminosity functions are 
similar, then the only effect is to adjust the comoving density estimates.  For example,
when we examined the effects of typing errors on the morphologically-typed luminosity
functions for the 2MASS survey, we found almost no effect due to random typing errors and 
only slow parameter drifts from typing biases (see \cite{kochanek00a}); this effect is
a direct result of the roughly constant numbers of galaxies of each morphological type 
near to the boundary between early- and late-type galaxies (S0/a).  The consequences
of both random and systematic errors are far greater when the shapes of the luminosity
functions are very different, as is found in the spectrally-typed surveys.
Figure 8 illustrates the problem for the Schechter function models of the high and low
equivalent width samples from Lin \etal\ (1996\markcite{lin96}).\footnote{These 
fits were done for the range $-23 < M < -17.5$ and
for this range they are consistent with non-parametric estimates of the luminosity
function.  The non-parametric luminosity function appears to be flatter than the
extrapolated Schechter function model for fainter galaxies (see \cite{lin96}).} 
Where the late-type galaxies dominate the comoving density, small ($f=5\%$!) 
classification problems for the late-type galaxies strongly modify the shape of the 
early-type luminosity function.  Even for purely random errors in the galaxy types,
Malmquist bias can mean that the observed density of the rarer galaxy type is 
dominated by misclassified galaxies of the common type.
 Large, luminosity dependent biases are needed to
make the early-type slope flatter and more similar to the late-type slope or the
results of the morphological surveys.  Crudely, a $f(L_*)=30\%$ error rate is 
enough to reduce the density of low EW galaxies near $L_*$ by a factor of two, 
which must be combined with a rapid reduction in the error for fainter galaxies so that
$f(L_*/10)\simeq 0$.  The corrected luminosity function which results from such
a bias has a flat faint-end slope.  Whether the actual biases can achieve this
is unknown.

\begin{figure}[t]
\figurenum{8}
\plotone{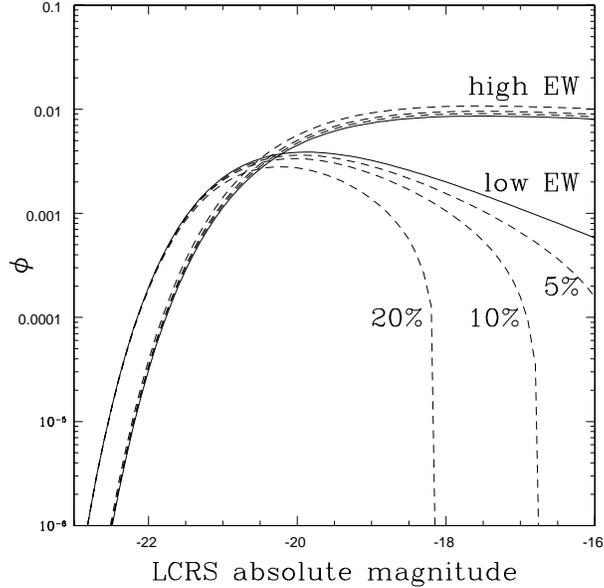}
\caption{Effect of unbiased misclassification on the LCRS luminosity functions
  divided by O[II]3727\AA\, equivalent width.  The solid curves are the parameterized
  luminosity functions for high ($>5$\AA) and low ($<5$\AA) EW from Lin \etal\ (1996).
  The dashed lines show the intrinsic luminosity functions assuming that $f=$5\%, 10\%,
  or 20\% of the intrinsically high EW galaxies are misclassified as low EW galaxies
  because of aperture bias independent of luminosity (the self-similar case).  
  Small misclassification rates lead to enormous changes in the intrinsic luminosity 
  function of the low EW galaxies because the high and low EW luminosity functions are 
  so different.  
\label{fig-lf-effects}
}
\end{figure}

\placefigure{fig-lf-effects}

The differences between the measured and global spectra affect not only the luminosity functions,
but also applications of the spectrally-typed surveys to other problems.
For example, Blanton (2000\markcite{blanton00}, also see \cite{tegmark99}) had already discovered inconsistencies 
in estimates of type-dependent cosmological biasing parameters from the LCRS survey when the sample 
was divided into high and low redshift samples.  Blanton (2000\markcite{blanton00}) found that surface brightness selection 
effects were probably not the origin of the problem, and suggested spectral typing problems as an alternative.  
Bromley \etal\ (1998b\markcite{bromley98b}) extended their earlier survey (\cite{bromley98a}) to estimate the luminosity
function by clan in high and low density environments.  They found that the luminosity functions
of the clans most sensitive to the aperture biases (clans 1--4) show stronger dependences on 
environment than those which are less sensitive (clans 5--6).  The sense of the difference, that the 
early-type clans in the low density, late-type rich environment showed still larger faint-end slope 
differences than the sample as a whole, is exactly the effect expected from typing biases in which spectrally
late-type galaxies are systematically misclassified as early-type galaxies.  Where the late-type
galaxy fraction is larger, the bias should be larger.  The aperture sizes are also small enough
to affect estimates of global star formation rates based on the emission lines at all redshifts
in the survey.  These may affect studies of the differences in star formation rates in different 
environments (e.g., \cite{hashimoto98}, \cite{allam99}) because the classification error rate
depends on environment. These particular studies minimized any biases by comparing galaxies with 
similar central concentrations to their luminosity profiles.

\begin{figure}[t]
\figurenum{9}
\plotone{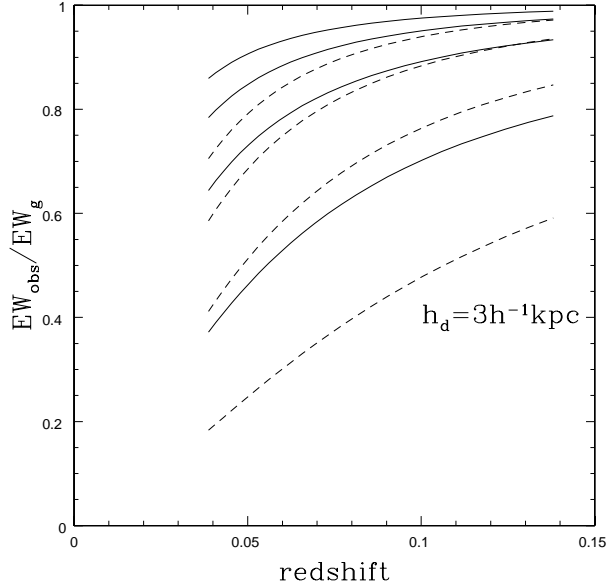}
\caption{
  The ratio of the measured to the global emission line equivalent width for galaxies
  with bulge-to-disk ratios of 10\% (solid) and 30\% (dashed) viewed through effective
  aperture diameters of (bottom to top) 3\farcs5, 7\farcs0, 10\farcs5, and 14\farcs0.  The scalings
  are the same as in Figures 6 and 7.
\label{fig-eqwidth-global2meas}
      }
\end{figure}

\placefigure{fig-eqwidth-global2meas}

\section{What Can Be Done?}

The lure of spectral typing---that it supplies a fast, quantitative method for dividing galaxy
samples into types---proves to be chimerical in its practical application because the precisely 
measured spectral types have an unknown, but inconsistent and biased, relation to the true 
spectral types.  We predict that the problems demonstrated in \S2 for the LCRS and ESP surveys
will also be found in the 2dFGRS and SDSS surveys, since the fiber apertures and mean redshifts
of the four surveys are all comparable.  Morphological classification appears to avoid such biases
for nearby galaxy samples (CfA, SSRS2, 2MASS), but is difficult to extend to the redshift range
or sample sizes of the deeper surveys.  In this section we discuss methods for correcting 
current spectral typing systems and alternative typing strategies.

The simplest solution to the problem would be to measure the global spectra of the galaxies.
For fiber spectrographs this is easy to implement by dithering the position of the instrumental
guide stars rather than holding them fixed during the integration.  The required dither amplitude
is surprisingly large.  Figure 9 shows the ratio between the true and global equivalent 
widths (as in Figs. 6 and 7) for effective aperture diameters of 3\farcs5 (the LCRS aperture), 
7\farcs0, 10\farcs5 and 14\farcs0.  Apertures two to three times larger than the 
LCRS aperture (i.e. comparable in size to the APM aperture) are needed to minimize the 
effects of aperture bias on spectral types.  Unfortunately, longer integrations are 
required because the continuum signal-to-noise ratio of the observations scales as the 
mean surface brightness of the galaxy inside the dithered aperture (assuming the noise
is dominated by the sky brightness), and for typical fiber sizes, the integration time for 
fixed signal-to-noise initially scales roughly as the aperture
diameter.  The signal-to-noise ratio for emission lines may actually be larger with dithering 
because the equivalent width of the emission lines rises with the increase in aperture size. 

If measuring the global spectra of all galaxies is impossible, global spectra can be measured
for a random sub-sample of the galaxies to build a statistical model for the effects of
aperture bias.  Enough spectra are required to build a contingency table for converting
survey spectral types into global spectral types.  Since a three-dimensional table is 
required (the global spectral type as a function of the survey spectral type and 
the apparent magnitude), the global spectra of $\sim 10^3$ galaxies are needed to 
provide adequate statistics.  Nonetheless, for large redshift surveys with $10^5$-$10^6$
galaxies, the overhead required to measure $10^3$ global spectra is only 1-10\% of the
total survey even if the integration times are 10 times longer.  This sampling method 
also has the advantage that it can be applied retroactively to an existing survey.
 
Rather than obtaining global spectra, a survey can try to obtain ``isophotal'' spectra for 
survey fields that require multiple exposures (more objects than fibers).  The targets
are divided into isophotal radius bins and the effective aperture (the amount of 
dithering) is scaled by the isophotal radius.  Perfect isophotal spectra avoid
redshift and flux dependent changes in the spectral types of galaxies with the 
same intrinsic properties.  Thus, isophotal spectra provide self-consistent galaxy 
types, although there will still be luminosity and structure dependent differences between 
the isophotal and global spectra.  By choosing the isophotes and effective apertures 
so that they match the fiber size for the faintest galaxies, the overhead of obtaining 
``isophotal'' spectra is minimized because large dithers are required only for the
brighter, nearby galaxies.  Instead of the rapid redshift variations in the ratio
of the measured to global equivalent widths seen in Figs. 6, 7, and 9, the ratios
found for isophotal spectra would be nearly independent of redshift. 
   
Surveys based on modern multi-color CCD photometry (SDSS and, to a lesser extent, 2MASS)
should probably type galaxies based on either broad band colors or fits to the luminosity
profiles of the galaxies (e.g. concentration or bulge-to-disk ratio).  While colors lack 
the physical precision of spectra, it is relatively simple to avoid redshift-dependent changes 
in the type-boundaries.  Colors are not free of biases, as dust creates strong, inclination
dependent variations in the colors of late-type galaxies.  Of the surveys we discuss, 
only the SDSS survey can easily define galaxy types based on broad band colors. Most of the 
other surveys have no color information or have only low precision photographic colors (APM, 
CfA, ESP, LCRS, SSRS2, and 2dFGRS), while the infrared colors measured by the 2MASS survey 
have little sensitivity to galaxy type.  One-dimensional light profile fits are the remaining possibility 
for automatically assigning galaxy types.  These may be the most reliable means of typing
galaxies, but they are difficult to automate.  In particular, all superposed stars and galaxies must
be identified and masked before performing the necessary profile fitting.  Like morphological
typing, profile typing requires good resolution and high dynamic range images.

In summary, all galaxy classification methods suffer from systematic errors and biases.  
We are relatively familiar with those of morphological classification, but no method is 
immune to the problem.  In particular, the application of spectral typing to large
redshift surveys can be easily automated and ``objective'' while still introducing
significant systematic biases and uncertainties. Because of the shear size of modern 
redshift surveys, these systematic errors are more important than statistical errors
and thus require careful attention in both the planning and analysis of such surveys.

\acknowledgments
The authors thank B. Bromley, H. Lin, J. Loveday, R. Marzke, L. Tresse,
and E. Zucca for helping us obtain the data necessary for this survey.  We also thank B. Bromley,
D. Fabricant, J. Huchra, R. Kirshner, M. Kurtz, O. Lahav, R. Somerville, and E. Zucca for extensive discussion of spectral 
typing and for their comments on the paper.  We thank the LCRS collaboration for making their data so 
readily available.  This admirable ease of access (rather than any form of animus) lead us to use the 
LCRS survey for most of our examples.  This research was supported by the Smithsonian Institution.
M.A.P. was supported by Hubble Fellowship grant HF-01099.01-97A from STScI 
(which is operated by AURA under NASA contract NAS5-26555).

{}

\end{document}